\documentclass{article}
\usepackage{spconf,amsmath,graphicx}
\usepackage{xcolor}

\usepackage{blindtext}
\usepackage{multirow}
\usepackage{cleveref}
\usepackage{accents}

\usepackage{enumitem}
\usepackage{etoolbox}
\usepackage{graphicx}
\usepackage{makecell}
\usepackage{caption}
\usepackage{wrapfig}
\usepackage{setspace}
\usepackage{subcaption}
\usepackage{floatflt}
\usepackage{float}
\usepackage{setspace}

\renewrobustcmd{\bfseries}{\fontseries{b}\selectfont}
\renewrobustcmd{\boldmath}{}
\newrobustcmd{\B}{\bfseries}

\newsavebox\CBox

\makeatletter

\renewcommand{\section}{\@startsection
   {section}%
   {1}%
   {}%
   {-0.4\baselineskip}%
   {0.2\baselineskip}%
   {}}%

\newcommand{\unnumberedsection}[1]{
    \@startsection{section}{1}{2.7cm}{-0.4\baselineskip}{0.2\baselineskip}{\normalfont\footnotesize\bfseries}*{#1}
}

\renewcommand{\subsection}{\@startsection
  {subsection}%
  {2}%
  {}%
  {-0.1\baselineskip}%
  {0.1\baselineskip}%
  {}}%

\renewcommand{\subsubsection}{\@startsection
  {subsubsection}%
  {3}%
  {}%
  {-0.2\baselineskip}%
  {0.2\baselineskip}%
  {}}%

\Crefname{equation}{Eq.}{Eqs.}
\Crefname{figure}{Fig.}{Figs.}
\Crefname{tabular}{Tab.}{Tabs.}
\Crefname{table}{Tab.}{Tabs.}
\Crefname{section}{Sec.}{Sec.}

\DeclareMathOperator*{\argmax}{arg\,max}
\DeclareMathOperator*{\argmin}{arg\,min}

\newcommand\numberthis{\addtocounter{equation}{1}\tag{\theequation}}

\title{On the Relation between Internal Language Model and \\Sequence Discriminative Training for Neural Transducers}
\name{Zijian Yang$^*$\thanks{$^*$ denotes equal contribution}, Wei Zhou$^{*\dagger}$\thanks{$^{\dagger}$ work done while at RWTH Aachen University, now at Meta}, Ralf Schlüter, Hermann Ney
\vspace{-2mm}
}
\address{Human Language Technology and Pattern Recognition, Computer Science Department,\\
RWTH Aachen University, 52074 Aachen, Germany,\\ AppTek GmbH, 52062 Aachen, Germany
\vspace{-3mm}
}
\begin{document}
%
\maketitle
\begin{abstract}
Internal language model (ILM) subtraction has been widely applied to improve the performance of the RNN-Transducer with external language model (LM) fusion for speech recognition. In this work, we show that sequence discriminative training has a strong correlation with ILM subtraction from both theoretical and empirical points of view. 
Theoretically, we derive that the global optimum of maximum mutual information (MMI) training shares a similar formula as ILM subtraction. 
Empirically, we show that ILM subtraction and sequence discriminative training achieve similar effects across a wide range of experiments on Librispeech, including both MMI and minimum Bayes risk (MBR) criteria, as well as neural transducers and LMs of both full and limited context.
The benefit of ILM subtraction also becomes much smaller after sequence discriminative training. 
We also provide an in-depth study to show that sequence discriminative training has a minimal effect on the commonly used zero-encoder ILM estimation, but a joint effect on both encoder and prediction + joint network for posterior probability reshaping including both ILM and blank suppression.
\end{abstract}
\section{Introduction \& Related Work}
\label{sec:intro}
Recently, sequence-to-sequence (seq2seq) modeling methods have achieved significant success in the realm of automatic speech recognition (ASR) \cite{prabhavalkar23e2e}. The most famous models include attention-based encoder-decoder (AED) models \cite{bahdanau2016end, chan2016listen, Tske2020SingleHA}, connectionist temporal classification (CTC) \cite{graves2006connectionist} and recurrent neural network transducers (RNN-T) \cite{graves2012sequence, Guo2020EfficientMW}. Although such seq2seq model can be used as a standalone model during decoding, integrating an external language model (ELM) trained on large amounts of text data often enhances performance. Shallow Fusion (SF) \cite{gulcehre2015using} is a popular approach to integrate the ELM during decoding, which simply employs a log-linear combination with seq2seq models. 

With the direct modeling of label sequence posteriors, seq2seq models implicitly learn an internal LM (ILM). According to Bayes interpretation, it is suggested to remove this ILM when integrating with an ELM.
Previous studies \cite{variani2020hybrid, mcdermott2019density, zeineldeen2021investigating, meng2021internal, zhou2022language, zheng2022empirical} have demonstrated that ILM subtraction can further improve the performance for both in-domain and cross-domain tasks.

Sequence discriminative training criteria including maximum mutual information (MMI) \cite{bahl1986maximum} and minimum Bayes risk training (MBR) \cite{Guo2020EfficientMW, meng2021minimum, weng2019minimum} also show significant improvements for seq2seq models (\cite{Guo2020EfficientMW, weng2019minimum, michel2020early}). \cite{michel2020early} introduces MMI training with ELM as an early-stage LM integration in the training phase. In \cite{meng2021minimum}, MBR training with LM integration achieves better performance than not using LM during training. In \cite{zeineldeen2021investigating}, MMI serves as an ILM suppression method for improving LM integration during decoding. 

In this work, we further investigate the relation between ILM subtraction and sequence discriminative training. Theoretically, we show a similar effect of ILM subtraction and MMI training by deriving the global optimum of MMI criterion. Empirically, we perform a series of comparisons between ILM subtraction and sequence discriminative training across different settings. Experimental results on Librispeech demonstrate that sequence discriminative training shares similar effects as ILM subtraction. Furthermore, we perform an extensive study of the impact of sequence discriminative training. Experimental results show a joint effect on both encoder and prediction + joint network to reshape posterior output including both label distribution and blank.

\vspace{-1mm}
\section{RNN-Transducer}
\vspace{-1mm}
\label{sec:transducer}
In this work, we apply the strictly monotonic RNN-T \cite{Tripathi19} as our acoustic model.
Let $X$ denote the input speech utterance and $a_1^S$ denote the output subword sequence from a vocabulary $\mathcal{V}$. 
Let $h_1^T$ be the encoder output sequence and $y_1^T$ be the blank $\epsilon$-augmented alignment sequence where $y_t \in \{\epsilon\} \cup \mathcal{V}$. 
The posterior of the output sequence $a_1^S$ is defined as:\\
\scalebox{0.9}{\parbox{1.1\linewidth}{%
\begin{align*}
    \begin{split}
        P_\text{RNNT}(a_1^S|X) &= \sum_{y_1^T: \mathcal{B}(y_1^T)=a_1^S} P_\text{RNNT}(y_1^T|h_1^T)\\
        &= \sum_{y_1^T: \mathcal{B}(y_1^T)=a_1^S} \prod_{t=1}^T P_\text{RNNT}(y_t|\mathcal{B}(y_1^{t-1}), h_t)
    \end{split}
\end{align*}}}
Here $\mathcal{B}$ is the collapse function, which maps $y_1^T$ to $a_1^S$ by removing $\epsilon$.
A limited context dependency based on a $k_\text{th}$ order Markov assumption can also be introduced to simplify the model \cite{ zhou2021phoneme, Prabhavalkar2021LessIM, Ghodis20}:\\
\scalebox{0.9}{\parbox{1.1\linewidth}{%
\begin{align*}
\notag
    P_\text{RNNT}(y_t|\mathcal{B}(y_1^{t-1}), h_t) = P_\text{RNNT}(y_t|a_{s_{t-1}-k+1}^{s_{t-1}}, h_t)
\end{align*}}}
Here $k$ is the context size and $s_1^T$ is the position sequence with $0 \leq s_t \leq S$ indicating the position in $a_1^S$ where $y_t$ reaches.
Given training pairs $(X_m, {a_{m,}}_1^{S_m})$ , the RNN-T model can be trained on the sentence-level cross-entropy (CE): \\
\scalebox{0.9}{\parbox{1.1\linewidth}{%
\begin{align*}
    \mathcal{L}_{\text{CE}} &= -\frac{1}{M} \sum_{m=1}^M \log P_\text{RNNT}({a_{m,}}_1^{S_m}|X_m) \notag \\
    & = -\sum_X Pr(X) \sum_{S, a_1^S} Pr(a_1^S|X) \log P_\text{RNNT}(a_1^S|X)
\end{align*}}}
where $Pr(X)$ and $Pr(a_1^S|X)$ denote the empirical distribution of the training data. The global optimum of the RNN-T model is obtained when $\hat{P}_\text{RNNT}(a_1^S|X)=Pr(a_1^S|X)$. 

For recognition with an ELM, we have the shallow fusion (SF) based decision rule:\\
\scalebox{0.9}{\parbox{1.1\linewidth}{%
\begin{align*}
   X \rightarrow {a^*}_1^{S^*} = \argmax_{S, a_1^S} \big[ P_\text{RNNT}(a_1^S|X) \cdot P^{\lambda_1}_\text{ELM}(a_1^S) \big]
   \numberthis \label{eq:desicionLM}
\end{align*}}}
as well as the Bayesian interpretation with ILM subtraction:\\
\scalebox{0.9}{\parbox{1.1\linewidth}{%
\begin{align*}
   X \rightarrow {a^*}_1^{S^*} = \argmax_{S, a_1^S} \big[ P_\text{RNNT}(a_1^S|X) \cdot \frac{P^{\lambda_1}_\text{ELM}(a_1^S)}{P^{\lambda_2}_\text{ILM}(a_1^S)} \big]
   \numberthis \label{eq:decisionILM}
\end{align*}}}
Here $\lambda_1$ and $\lambda_2$ are scales for $P_\text{ELM}$ and $P_\text{ILM}$, respectively.
Theoretically, The ILM of the model is defined as:\\
\scalebox{0.9}{\parbox{1.1\linewidth}{%
\begin{align*}
   P_\text{ILM}(a_1^S) = \sum_X Pr(X)  P_\text{RNNT}(a_1^S|X)
\end{align*}}}\vspace{-1mm}
which is infeasible to compute in practice. In practice, the $P_\text{ILM}$ can be estimated from the RNN-T model \cite{variani2020hybrid} or a standalone LM, known as density ratio (DS) \cite{mcdermott2019density}. In this paper, we apply zero-encoder (${h'}_\text{zero}$) \cite{meng2021internal, zhou2022language}, mini-LSTM \cite{zeineldeen2021investigating, zhou2022language} and DS as ILM estimation methods.

\section{Sequence Discriminative Training}
\subsection{MMI Training with LM Integration}
\subsubsection{Training Criterion}
The MMI training objective is defined as follows: \\
\scalebox{0.8}{\parbox{1.2\linewidth}{%
\begin{align}
    \mathcal{L}_\text{MMI} = -\sum_X Pr(X) \sum_{S, a_1^S} Pr(a_1^S|X) \log P_\text{seq}(a_1^S|X) \notag
    \numberthis \label{eq:MMI}
\end{align}}}
where the $P_\text{seq}(a_1^S|X)$ is a re-normalized probability of the combined RNN-T and LM probability with scales. \\
\scalebox{0.8}{\parbox{1.2\linewidth}{%
\begin{align*}
    P_\text{seq}(a_1^S|X) &= \frac{P^\alpha_\text{RNNT}(a_1^S|X) \cdot P^\beta_\text{LM}(a_1^S)}{\sum_{S',{a'}_1^{S'}}P^\alpha_\text{RNNT}({a'}_1^{S'}|X) \cdot P^\beta_\text{LM}({a'}_1^{S'})} \\
    &= \frac{P^\alpha_\text{RNNT}(a_1^S|X) \cdot P^\beta_\text{LM}(a_1^S)}{Z(X)}
    \numberthis \label{eq:qseqdef}
\end{align*}}}
where $\alpha$ and $\beta$ are scales, and $Z(x)$ is a normalization term. During training, the parameters of the LM are fixed, and only the RNN-T model is updated. The denominator is a summation of posteriors over all possible label sequences, which is infeasible in practice. Therefore, some approximation approaches are applied for the hypothesis space generation, such as N-best-list \cite{meng2021minimum, weng2019minimum} and lattice-free methods \cite{yang2023lattice}. 

\subsubsection{Global Optimum}
\label{sec:mmiopt}
Similar to the CE case, the global optimum of MMI training is obtained when $\hat{P}_\text{seq}(a_1^S|X) = Pr(a_1^S|X)$.
According to \Cref{eq:qseqdef}, we can derive the global optimum of the RNN-T model posterior as: \\
\scalebox{0.9}{\parbox{1.1\linewidth}{%
\begin{align*}
   &\hat{P}_\text{RNNT}({a}_1^{S}|X) = \big(\hat{Z}(X)\frac{Pr(a_1^S|X) }{P^\beta_\text{LM}({a}_1^{S})} \big )^\frac{1}{\alpha} \numberthis \label{eq:optfirst}
\end{align*}}}
Additionally, we have the normalization constraint\\
\scalebox{0.9}{\parbox{1.1\linewidth}{%
\begin{align}
\sum_{S, a_1^S} \hat{P}_\text{RNNT}({a}_1^{S}|X) = 1 = \hat{Z}(X)^\frac{1}{\alpha} \sum_{S, a_1^S} \big(\frac{Pr(a_1^S|X) }{P^\beta_\text{LM}({a}_1^{S})} \big )^\frac{1}{\alpha} \notag \\
\Rightarrow \hat{Z}(X)^\frac{1}{\alpha} = \frac{1}{\sum_{S, a_1^S} \big(\frac{Pr(a_1^S|X) }{P^\beta_\text{LM}({a}_1^{S})} \big )^\frac{1}{\alpha}} \label{eq:constz}
\end{align}}}
which shows that $\hat{Z}(X)$ is a constant value w.r.t the label sequence.
Therefore, we can drop $Z(X)$ in \Cref{eq:optfirst} and obtain\\
\scalebox{0.9}{\parbox{1.1\linewidth}{
\begin{align}
    \hat{P}_\text{RNNT}({a}_1^{S}|X) \propto \big (\frac{Pr(a_1^S|X) }{P^\beta_\text{LM}({a}_1^{S})} \big )^\frac{1}{\alpha} \numberthis \label{eq:optrnntmmi}
\end{align}}}\\
which shows that the global optimum of the RNN-T model posterior is reshaped (or subtracted in $\log$ space) by the text prior modeled by the LM used in training.

In this case, substituting \Cref{eq:optrnntmmi} into \Cref{eq:desicionLM} directly yields \Cref{eq:decisionILM} with the global optimum of $\mathcal{L}_{\text{CE}}$, which indicates the correlation between MMI training and ILM subtraction. However, in practice, the global optimum is not guaranteed due to limited training data, interdependencies in model parameters, regularization, and approximations of the hypothesis space employed. Similar to the effect of ILM subtraction discussed in \cite{zhou2022language}, MMI training boosts label probabilities and reshapes the distribution over labels.
Because ${P}_\text{RNNT}({a}_1^{S}|X)$ is normalized,
boosting label probabilities also leads to suppressing blank probabilities. These two effects are verified in experimental results in \Cref{sec:effectblank}.


\subsection{MBR Training with LM Integration}
\subsubsection{Training Criterion}
The general MBR training objective is defined as follows:\\
\scalebox{0.8}{\parbox{1.2\linewidth}{
\begin{align*}
    \mathcal{L}_\text{MBR} = \sum_{X} Pr(X) \sum_{S, a_1^S} Pr(a_1^S|X) \sum_{\Tilde{S}, \Tilde{a}_1^{\Tilde{S}}} P_\text{seq}(\Tilde{a}_1^{\Tilde{S}}|X) \cdot \boldsymbol{R}(\Tilde{a}_1^{\Tilde{S}}, a_1^S)
\end{align*}}}
where $\boldsymbol{R}$ is the cost function. In this work, we use the edit distance as the cost function.

\subsubsection{Global Optimum}
By exchanging the summation in $\mathcal{L}_\text{MBR}$, we have\\
\scalebox{0.9}{\parbox{1.1\linewidth}{
\begin{align*}
    \mathcal{L}_\text{MBR} =  \sum_{X} Pr(X) \sum_{\Tilde{S}, \Tilde{a}_1^{\Tilde{S}}} P_\text{seq}(\Tilde{a}_1^{\Tilde{S}}|X) \sum_{S, a_1^S} Pr(a_1^S|X) \cdot \boldsymbol{R}(a_1^S,\Tilde{a}_1^{\Tilde{S}})
\end{align*}}
}
For any $X$, there exists an optimal $\accentset{\ast}{a}_1^{\accentset{\ast}{S}}$ that minimizes the Bayes risk:\\
\scalebox{0.9}{\parbox{1.1\linewidth}{
\begin{align*}
\accentset{\ast}{a}_1^{\accentset{\ast}{S}} = \argmin_{\Tilde{S}, \Tilde{a}_1^{\Tilde{S}}}  \sum_{S, a_1^S} Pr(a_1^S|X) \cdot \boldsymbol{R}(a_1^S,\Tilde{a}_1^{\Tilde{S}})
\end{align*}}}
The global optimum of MBR training is then obtained when\\

\scalebox{0.8}{\parbox{1.2\linewidth}{
\vspace{-0.3cm}
\begin{align*}
    \hat{P}_\text{seq}(\Tilde{a}_1^{\Tilde{S}}|X) = \left \{
     \begin{array}{cc}
        1 \quad&,  \Tilde{a}_1^{\Tilde{S}} = \accentset{\ast}{a}_1^{\accentset{\ast}{S}}  \\
        0  \quad&, \text{otherwise}
    \end{array}
    \right .
\end{align*}}}
Besides similar constraints and approximations mentioned in MMI training, the peaky distribution makes the global optimum of MBR training harder to obtain. 
Therefore, we empirically show the correlation and similar effects of MBR training and ILM subtraction in \Cref{sec:exp}.


\section{Experiments}
\label{sec:exp}

\subsection{Setup}
We conduct our experiments on 960h Librispeech (LBS) \cite{panayotov2015librispeech}. The architecture of the RNN-T model follows \cite{zhou2022efficient}. The encoder consists of 12 conformer layers for both context-1 and full-context transducers. For the prediction network, we use two feed-forward layers in the context-1 model and one LSTM layer in the full-context transducer model. Gammaton features \cite{schluter2007gammatone} are applied as input and acoustic data-driven subword modeling (ADSM) units \cite{zhou2021acoustic} are used as output label units. We use RASR2 \cite{zhou2023rasr2} to perform all recognition and jointly with RETURNN \cite{zeyer2018returnn} for training.

We run sequence discriminative training as a fine-tune phase for CE-trained models, which follows the training pipeline proposed in \cite{zhou2022efficient}.
For the N-best-list methods, the hypotheses are generated using the CE-trained baseline model with a full-context LSTM LM SF.
While the N-best hypotheses remain unchanged, we also try to replace this full-context LM with a bigram LM for the training objectives.
Here the bigram and full-context LMs share the same architecture as the prediction network of transducers.  
For the lattice-free MMI approach, we follow the method proposed in \cite{yang2023lattice}. For each time step, only the best 20 candidates are kept to avoid the memory issue.
A transformer subword-level LM is used for recognition for all setups.

\begin{table}[t!]
\caption{{\it WER {[\%]} of full-context subword transducer models trained with different criteria on LBS. Sequence discriminative training uses full-context/bigram LM. Recognition with ELM uses a subword transformer LM and different fusion methods. (reduce-$P_\epsilon$ see \Cref{sec:effectblank})}}
\scalebox{0.9}{\parbox{1\linewidth}{
\setlength{\tabcolsep}{0.2em}
\begin{tabular}{|l|c|l|c|c|c|c|}
\hline
\multicolumn{2}{|c|}{Model Training} & \multirow{2}{*}{Recognition} & \multicolumn{2}{c|}{dev} & \multicolumn{2}{c|}{test} \\ \cline{4-7} 
 Objective & LM & & clean & other & clean & other \\ \hline
\multirow{6}{*}{CE} & \multirow{6}{*}{no} & no LM  & 2.7 & 6.9 & 2.9 & 7.1 \\ \cline{3-7} 
& & ELM SF & 1.9 & 4.5 & 2.1 & 4.9 \\ \cline{3-7}
& & \quad+ reduce-$P_\epsilon$ & 1.9 & 4.2 & 2.0& 4.7 \\ \cline{3-7}
& & \quad+ DR ILM          &  1.9   & 4.3 &  2.0   &  4.7   \\ \cline{3-7} 
& & \quad+ mini-LSTM ILM   & 1.8 & 3.9 & 1.9 & 4.3 \\ \cline{3-7}
& & \quad+ ${h'}_\text{zero}$ ILM & 1.8 & 4.0 & 1.9  &  4.3     \\ 

\hline \hline

\multirow{4}{*}{\shortstack[l]{+ N-best\\ \quad MMI}} & \multirow{2}{*}{\shortstack[c]{full\\context}} 
                       & ELM SF          & 1.8 & 4.0 &  1.9 & 4.3 \\ \cline{3-7} 
                       & & \quad+ ${h'}_\text{zero}$ ILM   &1.8  & 3.8 &1.8   & 4.3  \\ \cline{2-7}
 & \multirow{2}{*}{\shortstack[c]{bigram}}
                       & ELM SF          & 1.9 &  4.1  &1.9 & 4.4 \\ \cline{3-7}
                       & & \quad+ ${h'}_\text{zero}$ ILM   &1.9  & 4.0  &1.9 & 4.4 \\

 \hline \hline
                       
\multirow{4}{*}{\shortstack[l]{+ N-best\\ \quad MBR}} & \multirow{2}{*}{\shortstack[c]{full\\context}}
                       & ELM SF        &1.8  & 4.0 &  1.9 & 4.3 \\ \cline{3-7} 
                       & & \quad+ ${h'}_\text{zero}$ ILM & 1.7 & 3.9 & 1.9  & 4.2  \\ \cline{2-7}
 & \multirow{2}{*}{\shortstack[c]{bigram}} 
                       & ELM SF        & 1.8 & 4.1    & 1.9  & 4.4\\ \cline{3-7} 
                       & & \quad+ ${h'}_\text{zero}$ ILM & 1.8 & 4.0    & 1.9  & 4.2\\ \hline
\end{tabular}}}
\vspace{-1mm}
\label{tab:fullamce}
\end{table}

\begin{table}[t!]
\caption{{\it WER {[\%]} of context-1 subword transducer models trained with different criteria on LBS. Sequence training uses full-context/bigram LM. Recognition with ELM uses a subword transformer LM and different fusion methods.}}
\scalebox{0.9}{\parbox{1.0\linewidth}{
\setlength{\tabcolsep}{0.36em}
\begin{tabular}{|l|c|l|c|c|c|c|}
\hline
\multicolumn{2}{|c|}{Model Training} & \multirow{2}{*}{Recognition} & \multicolumn{2}{c|}{dev} & \multicolumn{2}{c|}{test} \\ \cline{4-7} 
 Objective & LM & & clean & other & clean & other \\ \hline
\multirow{4}{*}{CE} & \multirow{4}{*}{no} & no LM  & 2.7 & 7.3 & 2.9 & 7.4 \\ \cline{3-7} 
& & ELM SF & 1.8 & 4.2 & 1.9 & 4.4 \\ \cline{3-7}
& & \quad+ DR ILM          &   1.7  & 3.9 &  1.9   &  4.2   \\ \cline{3-7} 
& & \quad+ ${h'}_\text{zero}$ ILM     &  1.7   & 3.9    &  1.9   &  4.3     \\ 

\hline \hline

\multirow{4}{*}{\shortstack[l]{+ N-best\\ \quad MMI}} & \multirow{2}{*}{\shortstack[c]{full\\context}} 
                       & ELM SF          &1.7  & 3.8 &  1.8 & 4.2 \\ \cline{3-7} 
                       & & \quad+ ${h'}_\text{zero}$ ILM   & 1.7 & 3.8 & 1.8  & 4.2  \\ \cline{2-7}
 & \multirow{2}{*}{\shortstack[c]{bigram}}
                       & ELM SF          &1.7  &  3.9  &1.8 & 4.3 \\ \cline{3-7}
                       & & \quad+ ${h'}_\text{zero}$ ILM   & 1.7 & 3.9   &1.8 & 4.2\\

 \hline \hline
\multirow{2}{*}{+ LF MMI} & \multirow{2}{*}{\shortstack[c]{bigram}} 
                       & ELM SF          &1.8  & 3.9 & 1.9  & 4.2 \\ \cline{3-7} 
                       & & \quad+ ${h'}_\text{zero}$ ILM   & 1.8 & 3.9 & 1.9  & 4.2  \\ \cline{2-7}

\hline \hline

\multirow{4}{*}{\shortstack[l]{+ N-best\\ \quad MBR}} & \multirow{2}{*}{\shortstack[c]{full\\context}}
                       & ELM SF        & 1.7 & 4.0 & 1.8 & 4.3 \\ \cline{3-7} 
                       & & \quad+ ${h'}_\text{zero}$ ILM & 1.7 & 3.8 & 1.8  & 4.2  \\ \cline{2-7}
 & \multirow{2}{*}{\shortstack[c]{bigram}} 
                       & ELM SF        &  1.7 & 4.0    & 1.9  & 4.4 \\ \cline{3-7} 
                       & & \quad+ ${h'}_\text{zero}$ ILM & 1.7 &  3.9   &1.8  & 4.2 \\ \hline
\end{tabular}}}
\vspace{-4mm}
\label{tab:biamce}
\end{table}

\subsection{Full-context Transducer}
Table \ref{tab:fullamce} shows the comparison between ILM subtraction and sequence discriminative training for the full context RNN-T model. As expected, for the baseline model trained with CE, methods utilizing ELM yield substantial improvements over the standalone RNN-T system, and ILM subtraction approaches further improve the performance compared to the SF approach. For the models trained with sequence discriminative training criteria, the performance of SF is already comparable to the CE base with ILM subtraction. Moreover, when applying ILM subtraction on MMI/MBR-trained models, the improvement is notably smaller compared to the improvement seen in the CE-trained model, which indicates that the ILM is suppressed during sequence discriminative training.

For both MMI and MBR criteria, using a bigram LM for training results in a small degradation in performance compared to using the full context LM. We assume this is because the bigram LM introduces an ILM suppression that is inconsistent with the one captured by the RNN-T model, which makes the training suboptimum.
Therefore, the performance is slightly worse than the baseline using ILM subtraction.

\subsection{Context-1 Transducer}
Similarly, Table \ref{tab:biamce} exhibits comparisons of ILM subtraction and sequence discriminative training for context-1 transducer.
With a weaker ILM, the CE base model shows a worse standalone performance than the full-context transducer, but a better performance for ELM SF.
The benefit of ILM subtraction becomes generally smaller, and the best results with ELM are on par with full-context transducer.
Various sequence discriminative training settings yield similar performance as the CE base model with ILM subtraction, which indicates the claimed correlation also for context-1 subword transducers.

\begin{table}[t!]
\caption{\it PPLs of $h'_\text{zero}$ ILMs extracted from full-context transducer models trained with different criteria; and WERs of applying these ILMs (w/ renorm-$\epsilon$) to the baseline CE model for ILM subtraction in recognition with transformer ELM.}
\setlength{\tabcolsep}{0.35em}
\scalebox{0.9}{\parbox{1.0\linewidth}{
\begin{tabular}{|c|c|c|cc|}
\hline
$h'_\text{zero}$ ILM & \multicolumn{2}{c|}{PPL (dev-clean-other)}  & \multicolumn{2}{c|}{WER [\%]}                   \\ \cline{2-5} 
(from model) &       w/ renorm-$\epsilon$               &    w/o renorm-$\epsilon$                          & \multicolumn{1}{c|}{dev-other} & test-other \\ \hline
CE                        & 23                   & 160                          & \multicolumn{1}{c|}{4.0}       &    4.3        \\ \hline
MMI                       & 23                   & 108                           & \multicolumn{1}{c|}{4.0}       &     4.3       \\ \hline
MBR                       & 23                   & \phantom{1}98                          & \multicolumn{1}{c|}{4.0}       &  4.3          \\ \hline
\end{tabular}}}
\vspace{-4mm}
\label{tab:ilmppl}
\end{table}

\subsection{Effect of Sequence Discriminative Training on ILM}
To perform further in-depth study of the effect of sequence training on ILM, we employ the full-context transducer model, which shows a more noticeable effect of ILM.

\subsubsection{Effect on Prediction + Joint Network}
\label{sec:effectblank}
We first check the effect on the prediction + joint network alone using the $h'_\text{zero}$ ILM estimation from models trained with CE, MMI and MBR criteria. 
With a single softmax output for labels and blank $\epsilon$, the $h'_\text{zero}$ ILM is estimated by a simple renormalization excluding blank $\epsilon$ (renorm-$\epsilon$) \cite{meng2021internal, zhou2022language}.
Surprisingly, all these ILMs give the same perplexity (PPL) on the LBS dev set, as shown in Table \ref{tab:ilmppl}.
Further applying these ILMs to the CE base model for ILM subtraction in recognition also gives the same performance.
This suggests that MMI/MBR training has only a minimal impact on the label distribution estimated with the $h'_\text{zero}$ ILM approach.

Additionally, we also show the PPL of these ILMs w/o renorm-$\epsilon$ in Table \ref{tab:ilmppl}.
Smaller PPLs of ILM extracted from MMI and MBR models are observed, which indicates relatively smaller blank logits after sequence discriminative training, leading to the suppression of blank probabilities. 
We further simulate this effect in decoding by linearly/exponentially reducing the blank probabilities with a subsequent renormalization over $\{\epsilon\} \cup \mathcal{V}$ (denoted as reduce-$P_\epsilon$).
The result is shown in Table \ref{tab:fullamce}, which is better than CE baseline SF, but worse than MMI/MBR SF.
This indicates that sequence discriminative training is not only suppressing blank probabilities, but also reshaping the label distribution.

\subsubsection{Effect on Encoder and Prediction + Joint Network}
To further investigate the effect on the encoder and prediction + joint network, we swap these NN components from models trained with the three criteria.
All results for ELM SF are shown in Table \ref{tab:encswap}.
Using the CE model's encoder and the MMI/MBR model's prediction + joint network already improves over the CE baseline.
Using the MMI/MBR model's encoder and the CE models' prediction + joint network achieves similar improvement.
Using the full network of MMI/MBR models gives the best performance, with around 11\% relative improvement over the CE baseline.
All these indicate that sequence discriminative training has a joint effect on both encoder and prediction + joint network for probability reshaping including both ILM and blank suppression.

\subsection{Discussion of ILM Estimation}
As sequence discriminative training has a joint effect on the full network for better LM integration, a natural question is that why simply applying encoder-free ILM subtraction, e.g. $h'_\text{zero}$ ILM, on the CE base model can achieve similar performance.
Based on the results in this work, we infer that ILM estimated from the transducer's prediction + joint network is also highly correlated with the ILM effect contributed by the encoder.
This may also explain why the $h'_\text{zero}$ ILM performs better than the DR ILM approach as shown in Table \ref{tab:fullamce}. 


\begin{table}[t!]
\caption{\it WER {[\%]} of full-context transducer using encoder and prediction + joint network from models trained with different criteria. All recognition is with transformer ELM SF.}
\scalebox{0.9}{\parbox{1.0\linewidth}{
\begin{tabular}{|c|c|c|c|}
\hline
Encoder & Pred + Joint Network & \multicolumn{2}{c|}{WER {[\%]}}                    \\ \cline{3-4} 
 (from model) &  (from model) & dev-other & test-other \\ \hline
\multirow{3}{*}{CE}                                  & CE                                  & \multicolumn{1}{c|}{4.5}       & 4.9        \\ \cline{2-4} 
                                                     & MBR                                 & \multicolumn{1}{c|}{4.2}       &   4.6         \\ \cline{2-4} 
                                                     & MMI                                 & \multicolumn{1}{c|}{4.1}       &   4.6         \\ \hline
\multirow{2}{*} {MMI}                                     & CE                                 & \multicolumn{1}{c|}{4.1}       & 4.4        \\ \cline{2-4}
                                                        & MMI                                 & \multicolumn{1}{c|}{4.0}       & 4.3        \\ \hline
\multirow{2}{*} {MBR}                                     & CE                                 & \multicolumn{1}{c|}{4.1}       & 4.4        \\ \cline{2-4} 
                                                        & MBR                                 & \multicolumn{1}{c|}{4.0}       & 4.3        \\ \hline
\end{tabular}}}
\vspace{-4mm}
\label{tab:encswap}
\end{table}

\section{Conclusion}
In this work, we showed the strong correlation between ILM subtraction and sequence discriminative training for subword-based neural transducers. 
Theoretically, we verified that the global optimum of MMI training shares a similar formula with the ILM subtraction methods during decoding. 
Empirically, we showed that sequence discriminative training and ILM subtraction achieve similar effect across a wide range of experiments on Librispeech, including MMI and MBR training criteria, as well as different context sizes for both neural transducers and language models applied in training. 
Moreover, with an in-depth study, we showed that sequence discriminative training has a minimal effect on the commonly used zero-encoder ILM estimation, but a joint effect on both encoder and prediction + joint network for posterior probability reshaping including both ILM and blank suppression.
\scriptsize
\unnumberedsection{ACKNOWLEDGMENTS}
\scriptsize

\selectfont
This work was partially supported by NeuroSys, which as part of the
initiative “Clusters4Future” is funded by the Federal Ministry of
Education and Research BMBF (03ZU1106DA), and by the project RESCALE
within the program \textit{AI Lighthouse Projects for the Environment,
Climate, Nature and Resources} funded by the Federal Ministry for the
Environment, Nature Conservation, Nuclear Safety and Consumer
Protection (BMUV), funding ID: 67KI32006A.

\let\normalsize\small\normalsize
\let\OLDthebibliography\thebibliography
\renewcommand\thebibliography[1]{
        \OLDthebibliography{#1}
        \setlength{\parskip}{-0.3pt}
        \setlength{\itemsep}{1pt plus 0.07ex}
}

\bibliographystyle{IEEEbib}
\bibliography{strings,refs}

\begin{thebibliography}{10}

\bibitem{prabhavalkar23e2e}
Rohit Prabhavalkar, Takaaki Hori, Tara~N. Sainath, Ralf Schl\"uter, and Shinji Watanabe,
\newblock ``{End-to-End Speech Recognition: A Survey},''
\newblock Mar. 2023,
\newblock arXiv:2303.03329.

\bibitem{bahdanau2016end}
Dzmitry Bahdanau, Jan Chorowski, Dmitriy Serdyuk, Philemon Brakel, and Yoshua Bengio,
\newblock ``{End-to-End Attention-based Large Vocabulary Speech Recognition},''
\newblock in {\em ICASSP}. IEEE, 2016, pp. 4945--4949.

\bibitem{chan2016listen}
William Chan, Navdeep Jaitly, Quoc Le, and Oriol Vinyals,
\newblock ``{Listen, attend and spell: A neural network for large vocabulary conversational speech recognition},''
\newblock in {\em ICASSP}. IEEE, 2016, pp. 4960--4964.

\bibitem{Tske2020SingleHA}
Zolt{\'a}n T{\"u}ske, George Saon, Kartik Audhkhasi, and Brian Kingsbury,
\newblock ``{Single Headed Attention Based Sequence-to-Sequence Model for State-of-the-Art Results on Switchboard-300},''
\newblock in {\em INTERSPEECH}, 2020.

\bibitem{graves2006connectionist}
Alex Graves, Santiago Fern{\'a}ndez, Faustino Gomez, and J{\"u}rgen Schmidhuber,
\newblock ``{Connectionist Temporal Classification: Labelling Unsegmented Sequence Data with Recurrent Neural Networks},''
\newblock in {\em Proceedings of the 23rd international conference on Machine learning}, 2006, pp. 369--376.

\bibitem{graves2012sequence}
Alex Graves,
\newblock ``{Sequence Transduction with Recurrent Neural Networks},''
\newblock {\em arXiv preprint arXiv:1211.3711}, 2012.

\bibitem{Guo2020EfficientMW}
Jinxi Guo, Gautam Tiwari, Jasha Droppo, Maarten~Van Segbroeck, Che-Wei Huang, Andreas Stolcke, and Roland Maas,
\newblock ``{Efficient Minimum Word Error Rate Training of RNN-Transducer for End-to-End Speech Recognition},''
\newblock in {\em INTERSPEECH}, 2020.

\bibitem{gulcehre2015using}
Caglar Gulcehre, Orhan Firat, Kelvin Xu, Kyunghyun Cho, Loic Barrault, Huei-Chi Lin, Fethi Bougares, Holger Schwenk, and Yoshua Bengio,
\newblock ``{On Using Monolingual Corpora in Neural Machine Translation},''
\newblock {\em arXiv preprint arXiv:1503.03535}, 2015.

\bibitem{variani2020hybrid}
Ehsan Variani, David Rybach, Cyril Allauzen, and Michael Riley,
\newblock ``{Hybrid Autoregressive Transducer (HAT)},''
\newblock in {\em ICASSP}. IEEE, 2020, pp. 6139--6143.

\bibitem{mcdermott2019density}
Erik McDermott, Hasim Sak, and Ehsan Variani,
\newblock ``{A Density Ratio Approach to Language Model Fusion in End-to-End Automatic Speech Recognition},''
\newblock in {\em IEEE Automatic Speech Recognition and Understanding Workshop (ASRU)}. IEEE, 2019, pp. 434--441.

\bibitem{zeineldeen2021investigating}
Mohammad Zeineldeen, Aleksandr Glushko, Wilfried Michel, Albert Zeyer, Ralf Schl\"uter, and Hermann Ney,
\newblock ``{Investigating Methods to Improve Language Model Integration for Attention-based Encoder-Decoder ASR Models},''
\newblock in {\em INTERSPEECH}, 2021, pp. 2856--2860.

\bibitem{meng2021internal}
Zhong Meng, Naoyuki Kanda, Yashesh Gaur, Sarangarajan Parthasarathy, Eric Sun, Liang Lu, Xie Chen, Jinyu Li, and Yifan Gong,
\newblock ``{Internal Language Model Training for Domain-adaptive End-to-End Speech Recognition},''
\newblock in {\em ICASSP}. IEEE, 2021, pp. 7338--7342.

\bibitem{zhou2022language}
Wei Zhou, Zuoyun Zheng, Ralf Schl{\"u}ter, and Hermann Ney,
\newblock ``{On Language Model Integration for RNN Transducer based Speech Recognition},''
\newblock in {\em ICASSP}. IEEE, 2022, pp. 8407--8411.

\bibitem{zheng2022empirical}
Huahuan Zheng, keyu An, Zhijian Ou, Chen Huang, Ke~Ding, and Guanglu Wan,
\newblock ``{An Empirical Study of Language Model Integration for Transducer based Speech Recognition},''
\newblock in {\em Proc. Interspeech 2022}, 2022, pp. 3904--3908.

\bibitem{bahl1986maximum}
Lalit Bahl, Peter Brown, Peter De~Souza, and Robert Mercer,
\newblock ``{Maximum Mutual Information Estimation of Hidden Markov Model Parameters for Speech Recognition},''
\newblock in {\em ICASSP}. IEEE, 1986, vol.~11, pp. 49--52.

\bibitem{meng2021minimum}
Zhong Meng, Yu~Wu, Naoyuki Kanda, Liang Lu, Xie Chen, Guoli Ye, Eric Sun, Jinyu Li, and Yifan Gong,
\newblock ``{Minimum Word Error Rate Training with Language Model Fusion for End-to-End Speech Recognition},''
\newblock in {\em INTERSPEECH}, 2021, pp. 2596--2600.

\bibitem{weng2019minimum}
Chao Weng, Chengzhu Yu, Jia Cui, Chunlei Zhang, and Dong Yu,
\newblock ``{Minimum Bayes Risk Training of RNN-Transducer for End-to-End Speech Recognition},''
\newblock in {\em INTERSPEECH}, 2020, pp. 966--970.

\bibitem{michel2020early}
Wilfried Michel, Ralf Schl\"uter, and Hermann Ney,
\newblock ``{Early Stage LM Integration Using Local and Global Log-Linear Combination},''
\newblock in {\em INTERSPEECH}, 2020, pp. 3605--3609.

\bibitem{Tripathi19}
Anshuman Tripathi, Han Lu, Hasim Sak, and Hagen Soltau,
\newblock ``{Monotonic Recurrent Neural Network Transducer and Decoding Strategies},''
\newblock in {\em IEEE Automatic Speech Recognition and Understanding Workshop (ASRU)}, 2019, pp. 944--948.

\bibitem{zhou2021phoneme}
Wei Zhou, Simon Berger, Ralf Schl{\"u}ter, and Hermann Ney,
\newblock ``{Phoneme based Neural Transducer for Large Vocabulary Speech Recognition},''
\newblock in {\em ICASSP}. IEEE, 2021, pp. 5644--5648.

\bibitem{Prabhavalkar2021LessIM}
Rohit Prabhavalkar, Yanzhang He, David Rybach, Sean Campbell, Arun Narayanan, Trevor Strohman, and Tara~N. Sainath,
\newblock ``{Less is More: Improved RNN-T Decoding Using Limited Label Context and Path Merging},''
\newblock {\em ICASSP 2021 - 2021 IEEE International Conference on Acoustics, Speech and Signal Processing (ICASSP)}, pp. 5659--5663, 2021.

\bibitem{Ghodis20}
Mohammadreza Ghodsi, Xiaofeng Liu, James Apfel, Rodrigo Cabrera, and Eugene Weinstein,
\newblock ``{Rnn-Transducer with Stateless Prediction Network},''
\newblock in {\em ICASSP}, 2020, pp. 7049--7053.

\bibitem{yang2023lattice}
Zijian Yang, Wei Zhou, Ralf Schl{\"u}ter, and Hermann Ney,
\newblock ``{Lattice-Free Sequence Discriminative Training for Phoneme-Based Neural Transducers},''
\newblock in {\em ICASSP}, 2023, pp. 1--5.

\bibitem{panayotov2015librispeech}
Vassil Panayotov, Guoguo Chen, Daniel Povey, and Sanjeev Khudanpur,
\newblock ``{Librispeech: an ASR Corpus based on public Domain Audio Books},''
\newblock in {\em ICASSP}. IEEE, 2015, pp. 5206--5210.

\bibitem{zhou2022efficient}
Wei Zhou, Wilfried Michel, Ralf Schl\"uter, and Hermann Ney,
\newblock ``{Efficient Training of Neural Transducer for Speech Recognition},''
\newblock in {\em INTERSPEECH}, 2022, pp. 2058--2062.

\bibitem{schluter2007gammatone}
Ralf Schluter, Ilja Bezrukov, Hermann Wagner, and Hermann Ney,
\newblock ``{Gammatone Features and Feature Combination for Large Vocabulary Speech Recognition},''
\newblock in {\em ICASSP}. IEEE, 2007, vol.~4, pp. IV--649.

\bibitem{zhou2021acoustic}
Wei Zhou, Mohammad Zeineldeen, Zuoyun Zheng, Ralf Schl\"uter, and Hermann Ney,
\newblock ``{Acoustic Data-Driven Subword Modeling for End-to-End Speech Recognition},''
\newblock in {\em INTERSPEECH}, 2021, pp. 2886--2890.

\bibitem{zhou2023rasr2}
Wei Zhou, Eugen Beck, Simon Berger, Ralf Schl\"uter, and Hermann Ney,
\newblock ``{RASR2: The RWTH ASR Toolkit for Generic Sequence-to-sequence Speech Recognition},''
\newblock in {\em INTERSPEECH}, 2023, pp. 4094--4098.

\bibitem{zeyer2018returnn}
Albert Zeyer, Tamer Alkhouli, and Hermann Ney,
\newblock ``{{RETURNN} as a Generic Flexible Neural Toolkit with Application to Translation and Speech Recognition},''
\newblock 2018, pp. 128--133.

\end{thebibliography}

\end{document}